\def\Reals{{\rm I\!\rm R}} % open-face R, for real number line.
\def\R{{\Reals}}
\def\Q{{\bf Q}}  % rationals (need open face)
\def\Z{{\bf Z}}  % integers  (need open face)
\def\interior #1 { \buildrel\circ\over #1} % seems to work

\documentstyle[12pt]{article}
\begin{document}
\begin {titlepage}

\vspace{2 in}
\begin{center}
{\bf {\LARGE Morse Theory and the  \\ Topology of  Configuration Space}}

  W.D.McGlinn $^{\dagger}$,  L.O'Raifeartaigh $^{\ddagger}$,

 S.Sen $^{\sharp}$,  R.D.Sorkin*

 {\sl

 $^{\dagger}$ Physics Department, University of Notre Dame, Notre Dame, IN
   46556, USA

 $^{\ddagger}$ Dublin Institute for Advanced Studies, 10 Burlington
Road, Dublin 4, Ireland

 $^{\sharp}$ School of Mathematics, Trinity College, Dublin 2, Ireland

*Physics Department, Syracuse University, Syracuse, NY 13244-1130, USA} and
Departamento de Gravitaci\'on y Teor\'\i a de Campos
    Instituto de Ciencias Nucleares, UNAM
    A. Postal 70-543
    M\'exico, D. F. 04510, M\'exico

\end{center}

\noindent{\bf Abstract:}
The first and second homology groups, $H_{1}$ and $H_{2}$, are computed for
configuration spaces of framed three-dimensional point-particles with
annihilation included, when up to two particles and an antiparticle are
present, the types of frames considered being $S^{2}$ and $SO(3)$.  Whereas
a recent calculation for two-dimensional particles used the Mayer-Vietoris
sequence, in the present work Morse Theory is used.
   By constructing a potential-function none of whose critical
indices is less than four, we find that (for coefficients in an
arbitrary field $K$) the homology groups $H_{1}$ and $H_{2}$ reduce to
those of the frame-space, $S^{2}$ or $SO(3)$ as the case may be.
  In the case of $SO(3)$-frames this result implies that $H_{1}$ (with
coefficients in $\Z_{2}$) is generated by the cycle corresponding to a
$2\pi$-rotation of the frame.  (This same cycle is homologous to the
exchange loop: the spin-statistics correlation.)  It also implies that
$H_{2}$ is trivial, which means that there does not exist a topologically
nontrivial Wess-Zumino term for $SO(3)$-frames (in contrast to the
two-dimensional case, where $SO(2)$-frames do possess such a term).
   In the case of $S^{2}$-frames (with coefficients in $\Reals$), we
conclude $H_{2}=\Reals$, the generator being in effect the frame space
itself.  This implies that for $S^2$-frames there does exist a Wess-Zumino
term, as indeed is needed for the possibility of half-integer spin and the
corresponding fermi statistics.
  Taken together, these results for $H_1$ and $H_2$ imply that our
configuration space ``admits spin-1/2'' for either choice of frame, meaning
that the spin-statistics theorem previously proved for this space is not
vacuous.
\end{titlepage}

\section{Introduction}
Results concerning spin-statistics correlations for extended objects such
as solitons, monopoles and vortices have shown $^{1}$that the axioms of
local relativistic quantum field theory are by no means necessary to
guarantee a spin-statistics theorem$^{2}$.  The question of what
general$^{2}$ assumptions are needed for a spin-statistics theorem has led
to the investigation$^{3,4}$ of the general topological properties of
systems of particles and extended objects, with the result that the
physical consequences of these topological properties are now better
understood.  The importance of pair-creation and annihilation is
clear.$^{4,5}$

 The topology  of configuration space plays a important role in quantum
theory.  The
quantum mechanical Hilbert space corresponding to
a classical configuration
space $C$ is in general best viewed as being the space of sections of a
vector bundle over $C$.  For a variety of physically interesting
configuration spaces $C$, these vector bundles incorporate the spin-type
and the statistics, as well as other topological properties of the
quantum theory in question.  Furthermore many of the relevant
 topological properties can be described by the homotopy and homology
elements of the classical configuration spaces which are associated
with these vector bundles.

For example, the set of U(1) (and hence line) bundles over $C$ is
classified by $H^2(C;Z)$ , which is isomorphic to
$H_2(C;Z)^{*}\bigoplus Tor H_{1}(C;Z)$, where $H_{2}(C;Z)^{*}$ can
be thought of as the non-torsion part of $H_{2}(C;Z)$ and
$Tor H_1(C;Z)$ is the torsion subgroup of $H_{1}(C;Z)$.$^{6}$  (The torsion
subgroup of an abelian group is its maximal finite subgroup.)
Also, in the case of (locally) flat bundles, the
spin-statistics correlation follows from the statement that the
exchange of two identical particles and a $2\pi$ rotation of one of
the particles both correspond to the same nontrivial element of the
fundamental group $\pi_{1}(C)$.  For more general bundles it rests
(in three dimensions) on a homotopy, not between loops but between a
certain pair of mappings of $\Reals P^{2}$ into $C$.$^{7}$  Similarly, the
condition for the existence of a nontrivial Wess-Zumino term is that
the second homology group $H_{2}(C;Z)$ contain a Z subgroup, or
 equivalently that $H_2(C;\Reals)$ be non-zero.

A  system that has been investigated$^{8,9}$ using this type
of approach is that of identical particles and anti-particles on
$\R^d$, each carrying a `frame' F, the frame having been introduced in
order to describe intrinsic spin.$^{8,9}$  In this case, the classical
configuration space $C$ is (as a set) of the form $C = \bigsqcup
Q_{m,n}$ where the $Q_{m,n}$ are  spaces containing $m$ particles and
$n$ anti-particles, all with distinct locations (but see Ref. 10), and the
disjoint union
$\bigsqcup$ runs over all
possibilities for $m$ and $n$.  The topology of each subspace
$Q_{m,n}$ (or just $Q_{n }$ if the particle is its own antiparticle)
is the topology of an appropriate frame bundle modified by the fact
that the particles are assumed to be indistinguishable.  The basic
problem (solved in  Ref.~9) is to construct a
(Hausdorff) topology for the full space $C$ such that pair creation and
annihilation can proceed continuously.  The
problems of finding the precise topological properties that are
introduced by the construction of $C$ and of analyzing these properties
remains.

 In Ref.~11  we considered a limited version of this problem wherein
the individual units are point particles which move in two-dimensional
Euclidean space $\R^{2}$ and carry `frames' which embody
the notion of spin.  By `carrying frames' is meant that a single
particle (or anti-particle) is represented by a bundle over $\R^{2}$.
Three possible fibers, namely $SO(2)$, $ S^{2}$, and $SO(3)$, were
considered.  The restriction to point particles and to two dimensions
was for simplicity and because of the present interest in
two-dimensional systems, particularly in the theory of ``anyons''. In
this paper we expand the earlier study to include particles
which move in
three-dimensional Euclidean space $\R^{3}$ and carry frames $S^{2}$
or $SO(3)$.

The solution to the problem of finding an appropriate topology for the
configuration space is reviewed in Section 2.  The essential idea is
to introduce open neighborhoods of the vacuum (and corresponding
neighborhoods of the non-vacuum configurations) that allow a particle and
an antiparticle to annihilate provided their positions and frames are
suitably aligned.
(For an isolated particle-antiparticle pair, their frames must be ``mirror
images'' of each other; if a further particle is nearby in space, we
require in addition that the two particles be on opposite sides of the
antiparticle with both pairs of frames matched, a situation which we call
`syzygy' in analogy with
planetary alignments).  The complete topology is then obtained from
these neighborhoods. There remains the problem of determining in
detail the properties of the resulting topological space, including
its homology and homotopy groups, especially insofar as they help
answer the question of how many inequivalent vector bundles the space
admits.

 In the present work,
we will again concentrate on the homology groups $H_1$ and $H_2$,
these providing complete information if we restrict ourselves to line
bundles.  In Ref. 11
we determined these groups for particles moving in $\R^{2}$ by using
the Mayer-Vietoris exact sequence.  This process was carried out for
the first and second homology groups in the case of the subspaces
 $X_{1,1}={\overline Q_{1,1}}=Q_{1,1}\cup Q_{0,0}$ and
 $X_{2,1}={\overline Q_{2,1}}=Q_{2,1}\cup Q_{1,0}$.
In this paper we solve for the first and second homology groups for
 particles moving in three-dimensional Euclidean space $\R^{3}$,
carrying frames $S^{2}$ or $SO(3)$.  A result of particular importance
we find is that for $SO(3)$-frames,
 $H_{1}(X_{2,1};Z_2) = Z_{2}$,
the non-zero element being the exchange, or equivalently the
$2\pi$-rotation of a single frame. (That these two 1-cycles are
homologous expresses the spin-statistics correlation.)  This implies
that the particles will be spinorial fermions for an appropriate
choice of line bundle, and thereby
demonstrates the nontriviality of our
framework (at
least up to $X_{2,1}$).  Similarly we find that
 $H_{2}(X_{2,1};\R) = \R$ for $S^{2}$-frames, showing non-triviality
in this, somewhat more general case as well [see Ref. 7].  In addition we
find that $H_{2}$ is trivial for $SO(3)$-frames.

The layout of the paper is as follows.  In Sec. 2 we describe the topology
of the spaces to be considered.  In Secs. 3 and 4, we discuss how Morse
Theory can be used to obtain information about homology groups.  A Morse
potential applicable to our spaces is described in Secs. 5 and 6.  In
Sec. 7 its critical points are found, and in Sec. 8 the critical indexes of
these points are determined and the implications for $H_{1}$ and $H_{2}$ are
drawn.  Sec.~9 contains a discussion of the directions that generalizations
of the results of this paper may take.

\section{Topology of the Space of Framed Point-\hfil\break
Particles and Antiparticles}

The topology of the space of framed point-particles and antiparticles
is described in detail in Ref.~9.  For completeness, we give a brief
description of this space, including the `reflection' and `syzygy'
conditions for annihilation.  Let
$ X = (x, F^{(x)})$, [respectively $\bar{X} = (\bar{x}, \bar{F}^{(\bar{x})})$]
denote the position and `frame-orientation' of a particle
[resp. antiparticle]. By $F^{(x)}$ we mean a generic `frame'
(in $S^{2}$, $SO(3)$ or $SO(2)$ as the case may be)
attached to the particle located at position $x$.  Then
\begin{eqnarray}
Q_{m,n}  & = &  \{[X^{1}, X^{2}, \cdots, X^{m}; \bar{X}^{1}, \bar{X}^{2},
  \cdots , \bar{X}^{n}] | x^{i}, \bar{x}^{j }\in \R^{d}; \nonumber \\
  &  &     x^{i} \neq x^{j}, \bar{x}^{i} \neq \bar{x}^{j} \; {\rm if}
       \; i \neq j;  \ x^i \neq \bar{x}^{j}\}
\end{eqnarray}
is the sector or ``stratum'' of our configuration space
describing $m$ particles and $n$ antiparticles.
Here, the bracket notation indicates that the order of the $X^i$ is without
significance (and similarly for the $\bar{X}^j$).
 We also
introduce the vacuum (``VAC'') by setting
\begin{equation}
Q_{0,0} = \{VAC\}
\end{equation}

Next the concept ``$\epsilon$-close'' is defined as follows:
(i) Particles X and Y are ``$ \epsilon $-close'' iff
$ | x - y | < \epsilon $ and $ d(F^{(x)} , F^{(y)}) < \epsilon /L $ (where
L is some
fixed length), and similarly for antiparticles $\bar{X}$ and $\bar{Y}$;
(ii) The particle X and antiparticle $ \bar X $ are``$\epsilon$-close''
iff
$$
   |x - \bar{y }| < \epsilon \quad {\rm and} \quad
   d(F^{(x)}, R(x - \bar{x}) \bar {F}^{(\bar {y})}) < |x - \bar {y}| /L.
$$
Here $ |x - \bar {y}| $ is the Euclidean distance between points $x$
and $ \bar {y}$,
$d(F^{(x)},F^{(y)})$ is the geodesic distance between $ F^{(x)}$ and
$F^{(y)}$ in the space of frames, and $ R(x - \bar {y}) \bar
{F}^{(\bar{y})} $ is the frame which results when the anti-frame $ \bar
{F}^{(\bar {y})} $  is reflected in the plane perpendicular to the vector
$ x - \bar {y} $.
This concept of ``$\epsilon$-close'' is used to define an
$\epsilon$-neighborhood in $ Q_{m,n} $ of a point in $ Q_{m,n} $ in
 the obvious way.

We further need to define when a point

\[
{\bf Y} = [ Y^{1},\cdots,Y^{m+p}; \bar{Y}^{1},\cdots,\bar{Y}^{n+p} ]\in
Q_{m+p,n+p} \;\;\; , \ p \geq 0
\]

\noindent is in an $\epsilon$-neighborhood of a point

\[
{\bf X}=[ X^{1},\cdots,X^{m,}; \bar {X}^{1},\cdots \bar {X}^{n} ] \in Q_{m,n},
\]

\noindent that is when $p$ particle-antiparticle pairs are `close to
annihilation'.  To this end we define a {\it viable labeling} of $ {\bf
Y} $ with respect to $ {\bf X} $ as one that satisfies the following:
        (i) $ Y^{i}  $ is $ \epsilon$-close to $ X^{i } $ for i = 1,...,$m$ ;
        (ii) $ \bar {Y}^{i} $ is $ \epsilon$-close to $ \bar{X}^{i}$ for i =
             1,...,$n$ ;
         (iii) $ Y^{i +m} $ is $ \epsilon$-close to  $ \bar{Y}^{i + n } $  for
i
              = 1,..$p$.
Also, we say that a triplet  $ X,Y,\bar{Z} $ is in syzygy if (and only if)

\begin{equation}
\frac{|x-\bar{z}|+|y-\bar{z}|}{L}    >    \left\{ \begin{array}{l}
| \widehat{x-\bar {z}}+ \widehat{y-\bar{z}}| \\
d(F^{(x)},R(x-\bar{z})\bar{F}^{(\bar{z})})\\
d(F^{(y)},R(y-\bar{z}) \bar{F}^{(\bar{z})})
\end{array}\right.
\end{equation}
% remark: the left { is unmatched, and should be

\noindent A similar definition of syzygy applies to the triplet  ${\bar
X}, {\bar Y}, Z$.  Here $\widehat{x-\bar{z}} $  indicates the unit vector in
the direction $ x-\bar{z} $.
Finally we say that $ {\bf Y} \in N_{\epsilon} [{\bf X}] $, that is,$ {\bf
Y}$ is
an element
of the $ \epsilon$-neighborhood of $ {\bf X} $, if
there exists a viable labeling of $ {\bf Y} $ with respect to $ {\bf X} $,
and for all such viable labelings, all suitable triplets are in syzygy.  By
suitable triplets we mean that at least one member of the triplet comes
from the set  $ \{ Y^{m+1},.\cdots,Y^{m+p};\bar{Y}^{n+1},
\cdots,\bar{Y}^{n+p} \}$,  that is one `new' particle or one `new'
antiparticle must be a member of the triplet.

In summary, the condition
that a particle-antiparticle pair be ``close to annihilation'' is
first of all that they be spatially close to each other and that their
frames nearly satisfy the reflection condition. In addition, if
another particle is nearby in space, the three must be in syzygy.
This means that the two particles are on opposite sides of the
antiparticle and both particles nearly satisfy the reflection
condition with the antiparticle.

\section{Morse Theory}

Morse Theory$^{12}$ relates the homology groups of a manifold $M$ to the
so-called critical indices of a suitable smooth function $V$ on $M$.
Critical points of $V$ are points $p$ at which the gradient of $V$
vanishes: $d V | _{p}= 0$.  The number of negative eigensigns of the
Hessian $d^{2}V|_{p}$ at such a $p$ is referred to as the
index of  $p$ (relative to $V$), and $V$ is said to be
nondegenerate if (nullity of $p) := dim \, M - rank \, d^{2}V_{p} = 0$.

Let us assume that each of the
 sub-spaces $M_{a} = \{ s \in M | V(s) \leq a \}$ is
compact.  If $p$ is the only critical point of $V$ in the range $V(p) -
\epsilon < V(p) <  V(p) + \epsilon$ and $p$ is nondegenerate of index
$\lambda$, Morse theory tells us that $M_{V(p) + \epsilon} \approx M_{V(p)
- \epsilon} \cup e_{\lambda}$ where $e_{\lambda}$ is a $\lambda$ - cell and
`$\approx$' denotes homotopy equivalence.  (Considering a Morse function as
a potential, it effectively
induces a retraction of $M_{V(p) + \epsilon}$ to $ M_{V(p)
- \epsilon} \cup e_{\lambda}$.)  Because attaching $e_\lambda$ must create
either a new cycle or a new boundary, this in turn implies that either
\begin{eqnarray}
1)\;dimH_{\lambda}(M_{V(p)+\epsilon};K)  & =
  & dimH_{\lambda}(M_{V(p)-\epsilon};K) + 1 \;\;\;\;or \nonumber \\
 2)\; dim H_{\lambda-1}(M_{V(p)+\epsilon};K)& =
    &dim H_{\lambda-1}(M_{V(p)-\epsilon};K) - 1
\end{eqnarray}
Here $K$ is an arbitrary field and $dimH_{\lambda}(M;K)$ is the dimension
of the $\lambda^{th}$ homology group of $M$ with coefficients in $K$
(i.e. its dimension as a vector space over $K$).  To discover whether
alternative 1 or 2 obtains requires a global analysis.

       If one exhibits a smooth function $V$ on a closed manifold $M$ for
which all the critical points in $M\backslash M_0$ are nondegenerate (hence
finite in number), then $M$ can be constructed by the successive
attachment of thickened $\lambda_i$-cells to $M_0$, where $\lambda_i$ is
the index of the critical point $p_i$.  In particular, according to (4),
 dim~$H_\lambda$ can
change in going from $M_0$ to $M$ only if $\lambda=\lambda_i$ or
$\lambda=\lambda_i -1$ for some $i$.  From this it follows that, if
$\lambda_i \ge 4$ for all $p_i$, then ${\rm dim} H_j(M;K) = {\rm dim}
H_j(M_0;K)$, $j=1,2$. It is this result that we will use.

Actually we will need results which generalize the above.  First, for
`natural' Morse functions defined on our configuration spaces the critical
points $p_{i}$ will comprise {\it critical submanifolds} $N_{i}$ which
reflect an overall rotation invariance, as well as certain other
symmetries.  We can still apply Morse theory if the Hessian
$d^{2}V_{p_{i}}$ is nondegenerate in the directions normal to $N_{i}$, and
one still defines the {\it index} of $N_i$ as the number of negative
eigensigns of that Hessian.  Second, the spaces we will consider are not
globally manifolds nor are they compact.  The generalizations needed will
be described in the next section.

\message{section 4}
\section{Sufficient Conditions for Applying \hfil\break
         Morse Theory }

We would like to have information about the homology groups of the
space $X_{2,1} := {\overline Q_{2,1}} = Q_{1,0}\cup Q_{2,1}$, which is
not a manifold at points of its lower-dimensional ``stratum'' $Q_{1,0}$.
Nevertheless, we can still use Morse theory to get information
on $dim H_{i}(X_{2,1};K) - dim H_{i}(Q_{1,0};K)$.
We have shown elsewhere $^{11}$ that thickening $Q_{1,0}$
into $Q_{2,1}$ to obtain the $\epsilon$-neighborhood
$\widetilde {Q}_{1,0} = N_\epsilon[Q_{1,0}]$ does not alter the homology
of $Q_{1,0}$; in fact $\widetilde {Q}_{1,0}$ retracts onto $Q_{1,0}$ and
is therefore homotopy equivalent to it.  Outside of $\widetilde{Q}_{1,0}$
we do have a manifold (a subset of the top-dimensional ``stratum''
$Q_{2,1}$), and can therefore hope to use Morse theory to reduce
the study of $X_{2,1}$ to that of $\widetilde{Q}_{1,0}$, and thereby to
$Q_{1,0}$ itself.

To that end, let $M=Q_{2,1}\backslash\widetilde{Q_{1,0}}$ with boundary
$\Sigma = \partial M = \partial \widetilde{Q_{1,0}}$; and let
there be given on $M$ a Riemannian metric $g_{ab}$ and a smooth positive
potential-function $V$.
As earlier, we write $M_a = V^{-1}([0,a])$.
A sufficient set of conditions to apply Morse theory to $X_{2,1}$ in the
indicated manner is the following.

\begin{enumerate}
 \item  $\widetilde {Q}_{1,0}$ retracts onto $Q_{1,0}$
         (or onto any subset thereof onto which $Q_{1,0}$ itself retracts);
 \item $\Sigma = \partial M$
        is a piece-wise smooth submanifold of $Q_{2,1}$;
 \item  The gradient flow $\xi ^{a} = - g^{a b}\partial_{b}V$ is inward
        everywhere on $\Sigma$
        (by inward we mean into $\widetilde Q_{1,0}$, which is outward from
         the point of view of $M$);
 \item  The critical submanifolds of $V$ are compact (and nondegenerate),
        and there are a finite number of them in each $M_{a}$;
 \item  The vector flow $\xi^a$ leads each $x \in M$
         either to a point of $\Sigma$ or to a critical submanifold of $V$;
 \item The potential $V$ increases without bound along any trajectory of
        the inverse vector flow  $ - \xi^a$ (unless the trajectory hits a
        critical point).
\end{enumerate}

  The argument that the above conditions
suffice follows the same steps as the corresponding chain of reasoning from
ordinary Morse theory, the main differences being first, that  $M$ is
non-compact, and second, that the starting surface $\Sigma$ from which $M$
gets
built up is not a level surface of $V$.
(With our Morse function, there will be equipotential
surfaces of large $V$ intersecting $\Sigma$,
essentially because of the influence
of the term $V_\epsilon$, which we will introduce in the next section for
the sake of satisfying conditions 5 and 6 above.)

To cope with the circumstance that $V$ is not constant on $\Sigma$, we
introduce\footnote
{Another approach here might be to use the flow associated with $\xi$ for a
``Morse-Smale analysis''.  This could offer a different way around the
difficulty that $\Sigma$ is not a level surface of $V$.}
a slightly modified notation, letting
\begin{eqnarray*}
     M_t & = & \Sigma \, \cup \, V^{-1}([0,t])  \\
         & = & \{ x \in M | x \in \Sigma \; \vee \; V(x) \le t \} \
\end{eqnarray*}
 and
\[
   \Sigma_t =  V^{-1}(t) \cup  \{x\in\Sigma | V(x)>t \}
\]
\noindent
Thus, $M_t$ comprises the $V < t$ subset of $M$ {\it together with} that
portion of $\Sigma$ through which the $V<t$ subset does not yet
``protrude''; and $\Sigma_t$ is what might be called ``the future boundary
of $M_t$'' (see Figure 1).
Note that since $V>0$ everywhere, $\Sigma_0$ = $\Sigma$ =
$\partial M$ (= $\partial \widetilde{Q}_{1,0}$).  (Notice also that $M_t$
is not strictly a manifold, but would be if we thickened $\Sigma$ slightly
``back into'' $\widetilde Q_{1,0}$.)  We will denote the critical values of
$V$, taken in increasing order, by $v_1, v_2,\ldots$, each of which we
assume to correspond to a single connected critical submanifold $N_i$ of
$V$.

As usual, we consider the sequence of subspaces $M_t$ with $t$ increasing
from 0 to $\infty$, and we argue first that nothing happens between
critical values, and second that in passing a critical value $v_i$ we
merely in effect attach a cell of dimension $\lambda_i$ to $M$, $\lambda_i$
being the index of the $i^{th}$ critical point, as in the discussion of the
previous section.  Actually, since we will be dealing with critical {\it
submanifolds} this last statement must be modified, but the only difference
is that now, instead of the homology of $M$ being altered in dimension
$\lambda$ or $\lambda -1$ according to equation (4), it can be altered in
higher dimensions as well.  (A more precise statement is that, in passing a
critical submanifold $N$ of index $\lambda$, it is as if $b_k$ cells had
been added in dimension $\lambda + k$, $k = 0, 1, 2 \ldots$, where $b_k$ is
the $k^{th}$ betti number of $N$ with coefficients in a certain
orientation-bundle over $N$ (see ref. 12).  For our purposes this makes no
difference, and in the following discussion, we will speak as if all the
critical manifolds were simply (non-degenerate) critical points.

To begin with, let us argue that $M_t \approx \Sigma_0$ for all $t < v_1$
($\approx$ denoting homotopy equivalence).  For $t \le 0$,
$M_t=\Sigma=\Sigma_0$ and our claim is trivially true.  So consider any
$t>0$ but still less than $v_1$.  In order to retract $M_t$ back to
$\Sigma$, let us introduce on $M_t$ the renormalized Morse flow,
\[
   {\hat \xi}^a(x) = \;
     {- g^{ab}(x)\partial_b V(x) \over g^{cd}(x)\partial_c V(x)\partial_d V(x)}
    \;\;
    {V(b(x))-V(a(x)) \over t},
\]
where, for each $x\in M_t$, $a(x)$ is the unique point of $\Sigma_0$ to
which $x$ flows via the gradient flow, and $b(x)$ is the unique point of
$\Sigma_t$ to which it flows via
the reversed gradient flow.  (In other
words, $a$ and $b$ are the intersections of $\Sigma$ and $\Sigma_t$ with
the gradient flow-line through $x$.)
Since there are no critical points within
$M_t$, ${\hat \xi}^a$ is nonzero  and smooth,
except possibly on $\Sigma \cap \{V=t\}$, where it vanishes.
Moreover, ${\hat\xi}^a$ has been normalized so that
${\hat\xi}^a(x) \partial_a V(x) = -(V(b(x))-V(a(x)))/t$,
which means that the parameter difference along ${\hat\xi}^a$ between
$\Sigma_t$ and $\Sigma_0=\Sigma$ has the constant value $t$.  Thus the flow
induced by ${\hat\xi}^a$
effects a retraction of $M_t$ back onto $\Sigma_0$, whence these two
spaces are homotopic as claimed.
(By retraction, we always mean deformation
retract.)

To handle the similarly uneventful transition from $t = v_i + \epsilon$ to
$t = v_{i+1}-\epsilon$ we  can argue in the same manner that
$M'$ := $M_{v_{i+1}-\epsilon} \backslash \interior M_{v_i+\epsilon}$
 = $M_{ v_{i+1}-\epsilon} \backslash \{ V < v_{i}+\epsilon \}$
retracts onto
                 $\Sigma' := \Sigma_{v_i+\epsilon}$
and therefore
that
                 $M_{ v_{i+1}-\epsilon}$
retracts onto
                 $M_{ v_{i}+\epsilon}$ as desired.
This
proceeds as before with $M'$ playing the role of $M$ and $\Sigma'$ the role
of $\Sigma$, it being clear in particular that $\xi$ has no zeroes in $M'$
and that each $x\in M'$ flows to some point of $\Sigma'$.  Similarly,
nothing of note occurs beyond the final critical value $v_N$ (if there is
one), since there, a homotopy based on the Morse flow itself, i.e. on the
vectorfield
${\hat\xi}^a = - g^{ab}\partial_b V / g^{cd}\partial_c V\partial_d V$,
can be used to retract $M$ back to $M_{v_N+\epsilon}$.

Finally, we must analyze what happens in {\it passing} a critical value,
say $v_i$ with associated critical point $x_i$.  The key observation is
that, since everything really happens in a neighborhood of $x_i$  (or more
generally in a neighborhood of the {\it compact} critical submanifold $N_i$),
the effect on $M_t$ is just as it is in the standard situation discussed in the
last section.  More precisely, it follows from the analysis which applies
in that situation (see e.g. ref. 12) that there exist a standard neighborhood
${\cal U}$ of $x_i$, and decompositions
$$
  M^+ \backslash
             {\buildrel\circ\over {{M}^{-}}}    %% {\interior {M^-}}
   = {\cal U} \sqcup {\cal U'}  \quad\quad\quad
                         (M^\pm := M_{v_i\pm\epsilon})
$$
  and
$$
   \Sigma^- = S \sqcup S'  \quad\quad\quad
                             (\Sigma^\pm := \Sigma_{v_i\pm\epsilon})
$$
such that

(i) the flow $\hat\xi$ retracts ${\cal U'}$ onto $S'$, just as above;

(ii) there exists the usual Morse-theoretic retraction of ${\cal U}$ onto
    $S \cup e_\lambda$ (i.e.
a space made by attaching a $\lambda$-cell to $S$); and

(iii) the two retractions agree on
   $ \overline {\cal U} \cap \overline {{\cal U}'}$.

\noindent
Then the combined retractions in $\cal U$ and $\cal U'$ homotope
$M^+ \backslash {\buildrel\circ\over {{M}^{-}}}$  %% {\interior M^-}$
to $\Sigma^- \cup e_\lambda$, hence they also retract $M^+$ onto
$M^- \cup e_\lambda$, as required.  (See figure 2.)

Applying this analysis to the case at hand, we conclude that when the
conditions enumerated above are
fulfilled, $\overline{Q_{2,1}}$ is in effect built up from $Q_{1,0}$ by
adding cells of dimensions not less than the smallest index
$\lambda_{min}$ of any critical submanifold of $V$.  It follows that
$H_i(X_{2,1};K) = H_i(Q_{1,0};K)$ for all $i\le\lambda_{min}-2$.

\message{section 5}
\section{A Morse Function for $Q_{2,1}$}

Consider the following potential.

\begin{equation}
V = \left( {\it \rho_{1}} + {{{{{\it x_{1}}}^2}}\over {10\,{\it \rho_{1}}}}
\right) \,\left( {\it \rho_{2}} + {{{{{\it x_{2}}}^2}}\over {10\,{\it
\rho_{2}}}} \right) \,\left( \frac{1}{16} + {{x + {\it x_{1}} + {\it x_{2}}}
\over{{{\left( {{{\it \rho_{1}}}^2} + {{{\it \rho_{2}}}^2} +
 2\,{\it \rho_{1}}\,{\it \rho_{2}}\,\left( 1 - x \right)
\right) }^{{3\over 4}}}}} \right)
\end{equation}
Here (with respect to an arbitrary labeling of the particles)
$\rho_{i} = |x_{i}- \bar {y} |$ is the distance between particle $i$ and the
antiparticle, $x_{i}= 1-cos(\theta_{i})$ with $\theta_{i}$  the angular
separation $ d(F^{(x_{i})},R(x_{i} - \bar{y}) \bar {F}^{(\bar {y})}) $,
and $x = 1- cos(\theta)$ with $ \theta $ the angle between
$ - \vec{\rho _{1}}$ and $ \vec{\rho _{2}}$ so that
$ \rho^{2} = \rho^{2}_{1} + \rho^{2}_{2} + 2 \rho_{1}\rho_{2} (1- x)$ is
the distance squared  between the two particles; and we have taken
$L=1$.

The potential is the product of three factors.  The first two factors drive
the two particles to satisfy the reflection requirement with the
antiparticle and to move toward annihilation with it, thus tending to
produce a flow of $Q_{2,1}$ toward $Q_{1,0}$, as we desire.  Nevertheless
there is no obvious guarantee that points in $Q_{2,1}$ are not in some
cases driven toward two types of ``internal boundaries'':
1) where one or both of the particles are near to the antiparticle
   but not close to annihilation or
2) where the two particles are close to each other but not close to the
   antiparticle.
Either occurrence would violate condition 5 of the previous section.  In
fact, however, the `internal boundary' of type 2 is not approached by the
flow induced by $V$ because the $1/\rho^{3/2}$ in the
second term of the third factor drives the
particles apart unless they are in syzygy.
Despite this, $V$ can still become small if at at least one of the
first two factors becomes small, but in that case at least one particle
would have to draw near to the antiparticle, and we would be near a
boundary of type 1 rather than type two.

In order to insure that an internal boundary of
type 1 is also not approached  we will  add  a `small' term
$V_\epsilon$ to the potential:
\begin{equation}
  V \rightarrow V + V_{\epsilon}\,\,,\,\,\,\,V_{\epsilon} =
  \frac{\epsilon^{n}}{\rho_{1}\rho_{2}}
\end{equation}
where $n$ is some positive integer, say 6, and $\epsilon$ is the $\epsilon$
of $N_{\epsilon}$.  This term will drive the particles away from the
antiparticle when they get very close to it (the coupling $\epsilon^n$
being very small), but on the other hand we will take $n$ large enough to
avoid any violation of condition 3 of the previous section.
The addition of $V_\epsilon$ also ensures that condition 6 is fulfilled in
the neighborhood of the ``internal boundary''.

The following analysis assumes that the trivial
translational degree of freedom has been removed by bringing the
anti-particle to the origin via a preliminary retraction mapping.
This will allow all the critical submanifolds to be compact.

We are now almost ready to apply Morse Theory, but first we must establish
that our potential and metric lead to a flow which is inward everywhere
along the boundary $\Sigma$.

\section {The Flow at the Boundary $\Sigma$}

In order to verify condition 3, we need to characterize  $\Sigma$,
the boundary of $\widetilde {Q}_{1,0}$.  In fact we use a slightly modified
definition of $\widetilde {Q_{1,0}}$, as given by the following  conditions
(recall, we have set $L=1$):
\begin{enumerate}
\item  $x_1 < \rho_1  < \epsilon$,
\item  $x_2 < \rho_2$,   \
       $x   < \rho_2$.
\end{enumerate}
More precisely the condition is that there exist a labeling for which 1 and
2 hold; and we will always use such a labeling.  Moreover we will always
choose it so that $\rho_1\le\rho_2$, as is clearly possible since reversal
of 1 and 2 will not invalidate the conditions when $\rho_2\le\rho_1$.  [In
changing $\widetilde Q_{1,0}$ in this way we have in effect introduced a
modified conception of neighborhood (modifying the condition for syzygy)
which could be used to define a modified topology for
$X_{2,1}={\overline Q_{2,1}}$.  The new definition would be very similar to
the old one for small angles but easier to handle algebraically.  Unless we
actually alter the topology in this way, we do not guarantee that our new
$\widetilde{Q_{1,0}}$ will be a true neighborhood of $Q_{1,0}$, but that
doesn't matter: all we really need is that condition 1 of Sec. 4 still
obtain for our new $\widetilde Q_{1,0}$, and it does.]
The boundary $\Sigma$ is obtained by replacing any of the four inequalities
comprising conditions 1 and 2 by an equality,
for example $x_{1} = \rho_{1}$.

% Note, finally, that we can assume that $\rho_2\geq\epsilon$ when
% discussing the portion of the boundary characterized by $\rho_1 = \epsilon$.

There is considerable freedom in choosing the metric $g_{ab}$ which enters
into the Morse flow, and in particular influences the direction of the flow
across the boundary.  We choose
\begin{equation}
  ds^2 = \frac{\rho_1^2}{a(x_{1})} dx_1^2
  +\frac{\rho_{2}^{2}}{a(x_{2} )} dx_{2}^{2} + \frac{(\rho_{1}
  \rho_{2})^{2}}{a(x)}dx^{2}+ d\rho_{1}^{2} + d\rho
  _{2}^{2},
\end{equation}
where $a(x)$ is chosen to vanish linearly at $x=0,\,2$, and to rise rapidly
to a value of, say, 20 away from $x=0,\,2$.  More specifically, we assume that
$a(x)\sim x$ [resp. $2-x$] for $x\to 0$ [resp. 2].
Now (7) is not actually a metric on $Q_{2,1}$, but only on
the 5-dimensional quotient space of the parameters $\theta$ (or $x$) and
$\rho$ (call this space $P$).  Nonetheless, one can always choose a metric
on $Q_{2,1}$ such that the flow computed on $P$ is compatible with that on
$Q_{2,1}$.  Since each point of $P$ corresponds to a compact submanifold of
$Q_{2,1}$, this is all we will need.  (The condition for compatibility is
that the inverse metric $G^{AB}$ on $Q_{2,1}$ go over to the inverse of the
metric (7) on $P$ under the natural projection of $Q_{2,1}$ onto $P$; in
symbols
$g^{ab}(y) = G^{AB}(x) \partial y^a/\partial x^A \, \partial
y^b/\partial x^B$,
where $y\in P$ and $x\in Q_{2,1}$.)
By having the denominators of the angle terms in (7) vanish
linearly
at $x _{i} = 0,\,2 $ we ensure that $G^{AB}$ is
nonsingular at $\theta_{i} = 0,\,\pi$.

We now compute $\vec{\partial V}\cdot\vec{n}$ where $\vec{n}$ is a
conveniently chosen outwardly-oriented (co-)vector normal to the boundary
$\Sigma$ of $\widetilde {Q}_{1,0}$.  (In expressing vectors as $n$-tuples,
we take the coordinates in the order $(x_1,x_2,x,\rho_1,\rho_2)$.)

\message{____________________________________________________________}
\message{    WHEN THE FOLLOWING ERROR OCCURS, JUST HIT RETURN}
\message{------------------------------------------------------------}

\begin{enumerate}
\item For that part of the boundary defined by $x_1 = \rho_1 < \epsilon$,
we have,
with $\vec{n} = (1,0,0,-1,0)$ = $\partial x_1 - \partial\rho_1$,
\begin {eqnarray}
\frac{\vec{\partial V}\cdot\vec{n}}{\rho_{2} + \frac{x_{2}^{2}}{10
\rho_{2}}}& = &
{{11 a}\over {10\,{{{\it \rho_{1}}}^2}}}\,\left( {{{\it \rho_{1}}}\over
{{{\left( {{{\it
\rho_{1}}}^2} + {{{\it \rho_{2}}}^2} +
2\,{\it \rho_{1}}\,{\it \rho_{2}}\,\left( 1 - x \right)  \right) }^{{3\over
4}}}}}\right) \nonumber \\
&  &  \nonumber \\
&  &   + {{ a}\over {5 \,{{{\it \rho_{1}}}^2}}}\,\left( {1\over {16}} +
{{{\it \rho_{1}} + x + {\it x_{2}}}\over {{{\left( {{{\it \rho_{1}}}^2} +
{{{\it
\rho_{2}}}^2} + 2\,{\it \rho_{1}}\,{\it \rho_{2}}\,\left( 1 - x \right)
\right) }^{{3\over4}}}}} \right)  \nonumber \\
%% [latex adds an } above    ...^...  here it seems]
&  &  \nonumber \\
&  &  +  \left( {{33\,{\it \rho_{1}}\,\left( 2\,{\it \rho_{1}} + 2\,{\it
\rho_{2}}\,\left( 1 -
x \right)  \right) \,\left( {\it \rho_{1}} + x + {\it x_{2}} \right) }\over
{40\,{{\left( {{{\it
\rho_{1}}}^2} + {{{\it \rho_{2}}}^2} +
  2\,{\it \rho_{1}}\,{\it \rho_{2}}\,\left( 1 - x \right)  \right)
}^{{7\over 4}}}}}\right) \nonumber \\
&  &  \nonumber \\
&  &    -    {{9}\over {10}}\,\left( {1\over {16}} + {{{\it \rho_{1}} + x
+ {\it
x_{2}}}\over {{{\left( {{{\it \rho_{1}}}^2} + {{{\it \rho_{2}}}^2} +
 2\,{\it \rho_{1}}\,{\it \rho_{2}}\,\left( 1 - x \right)  \right)
}^{{3\over 4}}}}} \right)
\end{eqnarray}

All terms are positive except the last.  Consider the sum of the second
(positive) term and
the fourth (negative) term , that is
$$
\left(\, \frac{a \left( \rho_{1} \right)}{ 5 \rho_{1}^{2}}\, - \frac{9}{10}
 \, \right)\,
\left({1\over
{16}} + {{{\it
\rho_{1}} + x + {\it x_{2}}}\over {{{\left( {{{\it \rho_{1}}}^2} + {{{\it
\rho_{2}}}^2}
+ 2\,{\it \rho_{1}}\,{\it \rho_{2}}\,\left( 1 - x \right)  \right)
}^{{3\over 4}}}}}
\right)
$$
Since $\rho_1 < \epsilon$, this is clearly positive
(assuming $\epsilon\ll 1$) because $a\sim\rho_{1}$ for small $ \rho_{1}$.
Thus the flow is inward if we ignore the effect of $V_\epsilon$.  However,
since $V_{\epsilon}$ has no angle dependence
${\partial V_{\epsilon}}\cdot\vec{n}> 0$ is also positive, which only makes
things better.

\item For that part of the boundary defined by $x_2 = \rho_2$, with $\vec{n}
= (0,1,0,0,-1) = \partial x_2 - \partial\rho_2$ , we obtain the same
expression as in the previous case with 1 and 2 interchanged.
%% $\rho_{2}$ interchanged with $\rho_{1}$.
If $\rho_2 < \epsilon$, we are thus reduced to case 1 just
treated, so we may as well assume that $\rho_2 \ge \epsilon$.  Then for
$\rho_{2}$ not near the value 2 (where $a = 0 $), it is easy to see that
again the sum of the second and last terms of equation 8 is positive, since
its first factor, $a(\rho_2) / 5\rho_2^2 - 9/10$, exceeds $1/10$ if
$a(\rho_2)=20$.  For $x_2=\rho_2$ very close to $2$, this factor does
become negative, but in that regime the angle flow turns off and we are left
with
$$
\frac{\vec{\partial V}\cdot\vec{n}}{\rho_{1} +
\frac{x_{1}^{2}}{10\,\rho_{1}}} \approx
{{33\,\left( 2 + x \right) }\over {2^{7\over 2}\, 5}} - {{9\,\left( {1\over
{16}}
+ {{2
+ x}\over {{2^{{3\over 2}}}}} \right) }\over {10}}
$$
(Recall here that $x_1\le\rho_1 < \epsilon \ll 1$, so that
we may make the approximation $\rho_{1} = x_{1} = 0$.)
This is seen to be positive for all values of $x$ from 0 to 2.  Again we
see that the flow is inward, since as before
$\vec{\partial V_{\epsilon}}\cdot\vec{n}> 0$.

\item  For $\rho_1 = \epsilon$,
with $\vec{n} = (0,0,0,1,0) = \partial\rho_1$,
\begin {eqnarray}
\frac{\vec{\partial V}\cdot\vec{n}}{\left(\rho_{2} + \frac{x_{2}^{2}}{10\,
\rho_{2}}\right) } & = & {{-3\,\left( 2\,{\it \rho_{1}} + 2\,{\it
\rho_{2}}\,\left( 1 - x
\right)  \right) \,\left( {\it \rho_{1}} + {{{{{\it x_{1}}}^2}}\over {10\,{\it
\rho_{1}}}}
\right) \,\left( x + {\it x_{1}} + {\it x_{2}} \right) }\over {4\,{{\left(
{{{\it
\rho_{1}}}^2}
+ {{{\it\rho_{2}}}^2} + 2\,{\it \rho_{1}}\,{\it \rho_{2}}\,\left( 1 - x
\right)
\right)}^{{7\over 4}}}}} +\nonumber \\
& & \nonumber \\
&  & \left( 1 - {{{{{\it x_{1}}}^2}}\over {10\,{{{\it \rho_{1}}}^2}}} \right)
\left(
{1\over {16}} + {{x + {\it x_{1}} + {\it x_{2}}}\over {{{\left( {{{\it
\rho_{1}}}^2}
+
{{{\it \rho_{2}}}^2} +  2{\it \rho_{1}}{\it \rho_{2}}\left( 1 - x \right)
\right)
}^{{3\over 4}}}}} \right)
\end{eqnarray}

The first term is negative and the second positive.  Both attain their
minimum when $x_{1}$
takes its greatest allowed value of $\rho_{1}$, and for this condition
\begin {eqnarray*}
\frac{\vec{\partial V}\cdot\vec{n}}{\left(\rho_{2} + \frac{x_{2}^{2}}{10
\rho_{2}}\right)
 } & = &  - {{33\,{\it \rho_{1}}\,\left( 2\,{\it \rho_{1}} + 2\,{\it
\rho_{2}}\,\left( 1 - x
\right)  \right) \,\left( {\it \rho_{1}} + x + {\it x_{2}} \right) }\over
{40\,{{\left( {{{\it
\rho_{1}}}^2} + {{{\it \rho_{2}}}^2} +
  2\,{\it \rho_{1}}\,{\it \rho_{2}}\,\left( 1 - x \right)  \right)
}^{{7\over 4}}}}}
\\
&  &  \\
&  &    +{{9}\over {10}}\,\left( {1\over {16}} + {{{\it \rho_{1}} + x + {\it
x_{2}}}\over {{{\left( {{{\it \rho_{1}}}^2} + {{{\it \rho_{2}}}^2} +
 2\,{\it \rho_{1}}\,{\it \rho_{2}}\,\left( 1 - x \right)  \right)
}^{{3\over 4}}}}}
\right) \\
&  &  \\
 & = & \left( \frac{36}{40} - \frac{66}{40} +
\frac{6}{40}\left(\frac{\rho_{2}}{\rho_{1}}\right) \left(6
\frac{\rho_{2}}{\rho_{1}}+ 1 - x \right)\right)\left( > 0\right) +
\frac{9}{160}
\end{eqnarray*}

\noindent But $\rho_{2}$ is greater than  or equal to $\rho_{1}$
with our labeling convention.
Using this and setting $x = 2$ we find
$ \vec{\partial V}\cdot\vec{n}/ \left(\rho_{2} + \frac{x_{2}^{2}}
{10\rho_{2}}\right)  \geq + 9/160$.
Thus the flow without $V_\epsilon$ is inward for this part of the
boundary.  For this part of the boundary $\vec{\partial
V_{\epsilon}}\cdot\vec{n}/ \left(\,\rho_{2}+
\frac{x_{2}^{2}}{10\rho_{2}}\right) < 0$ . However
the most negative value it assumes on this boundary is $- \epsilon^{n-4}$,
and $n$ can be chosen large enough so that this negative contribution is
dominated by the positive $ \frac {9}{160}\,$ .

\item  Finally for the portion of the boundary defined by $x =  \rho_2$, with
 $\vec{n}$  = $\partial x - \partial\rho_2$ = $(0,0,1,0,-1)$,
 the following obtains.
\begin{eqnarray}
\frac{\vec{\partial V}\cdot  \vec{n}}{\rho_{1} +
\frac{x_{1}^{2}}{10\rho_{1}}} & = &   \frac{a \,\left( \rho_{2}
+\frac{x_{2}^{2}}{10\,\rho_{2}}\,\right)}{\rho_{2}^{2} \, \rho_{1}^{2} \,
\left(\, \rho_{2}^{2} + 2\, \rho_{2} \,\rho_{1} - 2 \rho_{2}^{2}  \rho_{1}
\,+ \rho_{1}^{2}\,\right)^{\frac{3}{4}}} \nonumber \\
&  &  \nonumber \\
&  & + \frac{\, 3\,a  \, \rho_{2}  \, \rho_{1}  \,\left( \, \rho_{2} + x_{2}
+ x_{1} \,\right)}{ 2 \, \left(\, \rho_{2}^{2} + 2\, \rho_{2} \,\rho_{1}
\,- 2 \, \rho_{2}^{2} \, \rho_{1} \,+ \rho_{1}^{2}\,\right)^{\frac{7}{4}}}
\nonumber \\
&   & \nonumber  \\
&  &     +  \left(\frac{3\left(  \rho_{2} + \rho_{1} - \rho_{2} \rho_{1}
\right)\left(  \rho_{2} + \frac{x_{2}^{2}}{10  \rho_{2}  } \right) \left(
\rho_{2} + x_{2}  + x_{1} \right) }{2 \left(\rho_{2}^{2} + 2 \rho_{2}
\rho_{1} - 2 \, \rho_{2}^{2} \, \rho_{1} \,+
\rho_{1}^{2}\,\right)^{\frac{7}{4}}}\right)  \nonumber \\
&  &  \nonumber \\
&  &  -  \left( 1 - \frac{x_{1}^{2}}{ 10 \rho_{1}^{2}} \right)\left(
\frac{1}{16} + \frac{ \rho_{2} + x_{2} + x_{1} }{\left( \rho_{2}^{2} + 2
\rho_{2} \rho_{1} - 2  \rho_{2}^{2}  \rho_{1} +
\rho_{1}^{2}\right)^{\frac{3}{4}}}\right)
\end{eqnarray}

\noindent The last term is negative.  However it is easy to argue that away
from $x = \rho_2 = 2$ , where $ a = 0$, the first  term (positive)
dominates the last term (negative).  For  $x = \rho_2 = 2$  we have
\begin{eqnarray*}
\frac{\vec{\partial V}\cdot  \vec{n}}{F (\rho_{1} +
\frac{x_{1}^{2}}{10\,\rho_{1}})}
& = &
\frac{3 \, \left(\, 2 - \rho_{1}\,\right)\, \left(\,2  +
\frac{x_{2}^{2}}{20}\,\right)\, \left(\,2 + x_{1} + x_{2}\,\right)}{2\,
\left( \,4 -
4\,\rho_{1} + \rho_{1}^{2}\,\right)^{\frac{7}{4}}}\\
&  &  \\
&  &  - \left(\,1 - \frac{x_{2}^{2}}{40}\,\right)\, \left(\,\frac{1}{16} +
\frac{2 +
x_{1} + x_{2}}{\left(\, 4  - 4\, \rho_{1} + \rho_{2}^{2}
\,\right)^{\frac{3}{4}}}\right)
\end{eqnarray*}

\noindent  To investigate the positivity of this expression we can let
$\rho_{1}$ and  $x_{1} \rightarrow 0 $ , since both are less then $\epsilon$,
and obtain

\begin{eqnarray*}
\frac{\vec{\partial V}\cdot  \vec{n}}{\rho_{1} + \frac{x_{1}^{2}}{10 \,
\rho_{1}}}
& = &
\frac{3\, \left(\, 2 + x_{2}\, \right)\, \left( \, 2 +
\frac{x_{2}^{2}}{20}\,\right)}{2^{\frac{7}{2}}}\\
& &  \\
& &  - \left(\, 1 - \frac{x_{2}^{2}}{40}\, \right)\, \left( \frac{1}{16} +
\frac{2
+ x_{2}}{2^{\frac{3}{2}}}\right)
\end{eqnarray*}

\noindent  This has it's minimum value at $x_{2} = 0$ and it is positive.
And once again $\vec{\partial V_{\epsilon}}\cdot\vec{n}> 0$~.

\end{enumerate}

\noindent We conclude that the flow is inward on  all of  $\partial
N_{\epsilon}$.

\section{Critical Points of $V$}

We now determine the critical points of $V$ and their indexes.
Consider the derivatives of $V$ with respect to the three angles:

\begin{eqnarray*}
\frac{\partial V}{\partial \theta _{1}}& = &{{\left( {\it \rho_{1}} +
{{{{\left( 1 - \cos ({\it \theta_{1}}) \right)}^2}}\over {10{\it \rho_{1}}}}
\right) \,
\left( {\it \rho_{2}} + {{{{\left( 1 - \cos ({\it \theta_{2}})
\right) }^2}}\over { 10{\it \rho_{2}}}} \right) \,
\sin ({\it \theta_{1}})}\over
{{{\left( {{{\it \rho_{1}}}^2} + {{{\it \rho_{2}}}^2} +
2\,{\it \rho_{1}}\,{\it \rho_{2}}\,\cos (\theta) \right) }^
{{3\over 4}}}}} + \\
  &  &   \\
   &  & {{\left( 1 - \cos ({\it \theta_{1}}) \right) \,
\left( {\it \rho_{2}} + {{{{\left( 1 - \cos ({\it \theta_{2}})
\right) }^2}}\over {10{\it \rho_{2}}}} \right) \,
\left( {1\over 16} + {{3 - \cos (\theta) - \cos ({\it \theta_{1}}) -
 \cos ({\it \theta_{2}})}\over
{{{\left( {{{\it \rho_{1}}}^2} + {{{\it \rho_{2}}}^2} +
2\,{\it \rho_{1}}\,{\it \rho_{2}}\,\cos (\theta) \right) }^
{{1\over 4}}}}} \right) \,\sin ({\it \theta_{1}})}\over
{{5 \it \rho_{1}}}}\\
  &  &  \\
\frac{\partial V}{\partial \theta _{2}}& = & \frac{\partial V}{\partial
 \theta _{1}}(1 \leftrightarrow 2) \\
  &  &  \\
\frac{\partial V}{\partial \theta} & = & \left( {\it \rho_{1}} +
{{{{\left(1 - \cos ({\it \theta_{1}}) \right) }^2}}\over {10{\it \rho_{1}}}}
\right) \,
  \left( {\it \rho_{2}} +
{{{{\left( 1 - \cos ({\it \theta_{2}}) \right) }^2}}\over {10{\it \rho_{2}}}}
\right) \, \\
  &  &  \\
  &  & \left( {{\sin (\theta)}\over
{{{\left( {{{\it \rho_{1}}}^2} + {{{\it \rho_{2}}}^2} +
   2\,{\it \rho_{1}}\,{\it \rho_{2}}\,\cos (\theta) \right) }^{{3\over 4}}}}} +
{{{2 \it \rho_{1}}\,{\it \rho_{2}}\,
 \left( 3 - \cos (\theta) - \cos ({\it \theta_{1}}) -
 \cos ({\it \theta_{2}}) \right) \,\sin (\theta)}\over
{3\,{{\left( {{{\it \rho_{1}}}^2} + {{{\it \rho_{2}}}^2} +
2\,{\it \rho_{1}}\,{\it \rho_{2}}\,\cos (\theta) \right) }^{{7\over 4}}}}}
\right)
\\
 \end{eqnarray*}

\noindent
{}From these expressions it is clear that critical points can occur only
for all the $\theta$'s equal to 0 or $\pi$.  To find the critical points of
$V$ it thus suffices to find its critical points with respect to $\rho_{1}$
and $\rho_{2}$, assuming fixed $\theta$'s of $0$ or $\pi$.  To that end we
exhibit contour plots for $V$ with $\Theta=(\theta_{1},\theta_{2},\theta)$
taking these values.

Figure 3 is a contour plot for $V$ in terms of $\rho_{1}$ and $\rho_{2}$
for $\Theta = (\pi,\pi , 0 )$.  We see that there are three critical points,
one at $P_{1,1,0} = (\theta_{1}, \theta_{2}, \theta , \rho_{1}\ \rho_{2})
\approx (\pi,\pi,0, 1.5, 1.5)$ and a symmetrical pair, one partner of which
is  at
$P_{1,1,0}=(\theta_{1},\theta_{2},\theta,\rho_{1}\ \rho_{2})
\approx (\pi,\pi,0, 0.7, 8.0)$.

Figure 4 is a plot for $V$ for $\Theta = (\pi,\pi,\pi )$.  From this plot
we see there is a critical point at
$P_{1,1,1}\approx(\pi,\pi,\pi,16,0.6)$.

Similarly Figure 5 is a contour plot for $V$ with $\Theta=(0,\pi,\pi)$.
This plot exhibits a critical point at $P_{0,1,1} \approx
(0,\pi,\pi,12,0.6)$.

The plots for the other four possible values of $\Theta$ depict functions
with no critical points, and we do not give them here.

{}From these plots we see that, at these critical points, the eigensigns of
the Hessian restricted to the $\rho_{1}$-$\rho_{2}$-tangent-subspace are
positive except for the critical point $P_{1,1,0} =
(\theta_{1},\theta_{2},\theta,\rho_{1}\ \rho_{2}) \approx
(\pi,\pi,0,1.5,1.5)$ of Figure 3, which is a saddle point with one negative
eigensign and one positive one.

{}From the plots, it seems clear that the addition of the `small' term
$V_{\epsilon}$ (which has no angle dependence) will not introduce any new
critical points and will affect only slightly the position of the critical
points without changing their indexes. It is also clear from the above
expressions for the $\theta$-derivatives of $V$ that the $5\times 5$
Hessian matrix of $V$ at a critical point has no off-diagonal terms
involving $\theta$'s.  The geometries of the three configurations are
depicted in Figure 6.

 If one contemplates
the types of configuration implied by the values of $\Theta$ at these
critical points , it becomes plausible
that the signs of the diagonal $\theta$-entries should behave as
follows.
\begin{enumerate}
\item
$P_{1,1,0}$ should have two negative signs for $\theta_{1}$ and
$\theta_{2}$, since the particle frames are oppositely aligned to that
of the antiparticle (an unstable situation), and one positive sign for
$\theta$, since a change of $\theta$ in this case would bring the two
particles {\it out} of syzygy (a stable situation). Thus the
total critical index of
$P_{1,1,0}  \approx
(\pi,\pi,0, 1.5, 1.5)$ should be 3 while that of $P_{1,1,0}  \approx
(\pi,\pi,0, .7, 8)$ should be 2.
\item
$P_{1,1,1}$ should have three negative signs since the frames
are anti-aligned and their locations are in ``anti-syzygy'', and thus a
critical
index of 3

\item  $P_{0,1,1}$ should have two negative signs for $\theta_{2}$ and
$\theta$, due to frame misalignment and being out of syzygy, and one
positive sign for $\theta_{1}$, due to frame alignment, leading to a critical
index of 2
\end{enumerate}

These conclusions are readily confirmed by computing the Hessian of $V$
in the five dimensional space of $\theta_{1}, \theta_{2}, \theta,
\rho_{1}, \rho{_2}$ at these critical points.

\section{Critical Indexes of $V$}

Before turning to the case of primary interest, let us consider what
the Morse function $V$ tells us when the particles and antiparticle are
assumed to move in the Euclidean {\it plane} $\R^2$, and the
frame-space is correspondingly $SO(2)$.  In this case each of the
critical points lies on a
two-dimensional
critical submanifold N$_{i}$ generated by ({\it a}) overall rotation [$SO(2)$]
of the points and frames together, and ({\it b}) ``locked frame rotation''
[$SO(2)$] in which both of the frames rotate clockwise (say) while the
anti-frame rotates counterclockwise in order that $\theta_1$ and $\theta_2$
be preserved.
 The Hessian to be considered is that of the normal subspace to
N$_{i}$.
Thus the critical index  for $(\pi, \pi, 0, 1.5, 1.5)$ is 3, for $(\pi,
\pi, 0, .7, 8)$
2, for $(\pi, \pi, \pi, 16, .6)$   3 and for   $(0, \pi, \pi, 12, .6)$  2.
By equation
(4) we
 can conclude that (for any choice of the coefficient field $K$),
dim$H_1$ cannot increase in going from $Q_{1,0}$ to
$X_{2,1}=\overline{Q_{2,1}}$ but might decrease by as much as 2, while
dim$H_2$ might change by any integer in the range $[-2,+2]$.

 Recall now that $Q_{1,0}\approx\bar{F}$, where $\bar{F}$ is the
frame-space of the antiparticle.  Since we now know that no new
generator of $H_1$ can appear in going from $Q_{1,0}$ to $X_{2,1}$, we
can conclude that the exchange path in $X_{2,1}$ is either homologous
to the cycle in $Q_{1,0}$ corresponding to a $2\pi$ rotation, or
homologous to zero.  In Ref. 11 we in fact concluded that the former
obtains.  (The above conclusion on how dim$H_2$ changes is also consistent
with the findings of that reference.)

Now let us turn to the case of primary interest,
where the particles and antiparticles are assumed to
move  in Euclidean 3-space $\R^{3}$.  We can anticipate that the
 critical indexes
are considerably enlarged, because both the dimension of the physical
Euclidean
space and the dimension $d_{F}$ of the frame  space  have grown.  At
each of the critical points,
 the unstable directions amount to rotating the frames out of
anti-alignment  or the frame-locations out of anti-syzygy keeping the
antiparticle frame fixed.
The critical submanifolds now are respectively 4 or 5 dimensional (3
dimensions for overall rotations, plus $d_F$ dimensions for ``locked
frame rotations'' minus 1 dimension over-counted), and the critical
indexes refer -- as always -- to the directions normal to the critical
submanifolds.

Consider first the case of $P_{1,1,0}$: both particle frames
anti-aligned but with location-syzygy.  There are $2 d_{F}$ unstable
directions to rotate the frames of the particles, giving for the
critical indexes $\lambda_{1,1,0}^{(1)} = 2 d_{F}+1$   for $(\pi, \pi, 0, 1.5,
1.5)$
and $\lambda_{1,1,0}^{(2)} = 2 d_{F} $   for $(\pi, \pi, 0, .7, 8) $

In the case of $P_{1,1,1}$, both particle frames are anti-aligned and
there is location-anti-syzygy.  In addition to the $2 d_{F}$ unstable
directions to rotate the frames of the particles there are 2 unstable
directions arising from perturbing the locations.  (These two directions
correspond to the two independent ways that $\vec\theta $ can  change
when the particles can move in three directions.  This is in contrast
to the two dimensional case for which there is only one independent
way.)  Thus  $\lambda_{1,1,1} = 2 d_{F} + 2$.

Finally in the third case of $P_{0,1,1}$ (particles in
location-anti-syzygy with the closest particle anti-aligned), there
are $d_{F}$ unstable direction to rotate the misaligned frame and, as
in the $P_{1,1,1}$ case, there are 2 unstable directions in which to
alter the locations, giving $\lambda_{0,1,1} = d_{F} + 2 $.

Thus for  $S^{2}$ frames
$\{ \lambda_{1,1,0}^{(1)}, \lambda_{1,1,0}^{(2)},
\lambda_{1,1,1},\lambda_{0,1,1}\}
= \{5, 4, 6, 4\}$
whereas for $SO(3)$ frames
 $\{ \lambda_{1,1,0}^{(1)}, \lambda_{1,1,0}^{(2)},
\lambda_{1,1,1},\lambda_{0,1,1}\}
= \{7, 6, 8, 5\}$.
We see that overall, the minimum index is 4, which is too large to
influence $dimH_{1}(...;K)$ or $dimH_{2}(...;K)$ in going from $Q_{1,0}$ to
$X_{2,1}$; and we conclude that, for both types of frame,
\begin{equation}
dimH_{i}(X_{2,1};K) = dimH_{i}(Q_{1,0};K) = dimH_{i}( F;K) ,   \; \; \; i
= 1,2.
\end{equation}

It bears emphasis here, that this conclusion has turned out to be
independent of the details of the $\rho_1$--$\rho_2$ plots given in figures
3--5.  In fact the only type of critical point which did not occur in these
plots was one with two negative eigensigns, but such a situation would only
have increased the resulting critical indices found above, and therfore
would not have disturbed our main conclusion, equation (11).  In retrospect
we can see that this conclusion follows directly from the observation made
earlier, that for our Morse potential, critical points can occur only for
all the $\theta$'s equal to 0 or $\pi$.  In particular any critical points
in the above plots which might have been overlooked (including ones
conceivably introduced by our addition of the $V_\epsilon$ term), would
have been harmless anyway.

Equation (11) is true for any coefficient-field $K$.  We will need it for
$K=\R$ (or equivalently $\Q$) and $K=\Z_2$, the two-element field.
Let us take the cases $F=SO(3)$ and $F=S^2$ in that order.

\smallskip
\noindent
1) $F = SO(3)$.   We know that
 $H_{1}(SO(3);\Z_{2}) = \Z_{2} $, the generator being the 1-cycle
corresponding to 2$\pi$-rotation of the frame.  Thus,
\[
     dim H_{1}(X_{2,1};\Z_2) = dim H_{1}(SO(3);\Z_{2}) = 1.
\]
We can conclude that the rotation remains nontrivial in
$X_{2,1}$, and that the exchange one-cycle (which also generates a
$Z_{2}$ if it is nontrivial) is either homologous to a frame-rotation,
or to zero.  By the spin-statistics theorem of Ref. 9 we know that
the former  obtains.  Most importantly we conclude that since
neither of these cycles is homologous to zero in $X_{2,1}$, this
configuration space does in fact ``admit spin 1/2'', as is needed to
avoid our entire framework being essentially vacuous as regards
questions of spin and statistics.  Further since
$H_{2}(SO(3);\Reals)=0$ and thus
\[
   0 = dimH_{2}(SO(3);\Reals)  = dimH_{2}(X_{2,1};\Reals),
\]
there exists no nontrivial Wess-Zumino term for $X_{2,1}$.

\smallskip
\noindent
2) $F = S^{2}$.  In this case, $H_{1}(S^{2};Z_{2}) = 0$, whence
\[
   0 = dimH_{1}(S^2;Z_{2})  = dimH_{1}(X_{2,1};Z_{2}).
\]
We conclude that in $X_{2,1}$ the exchange one cycle must also be
homologous to zero, i.e. that  rotation and exchange are both trivial in
$H_1$.  However for $S^2$-frames, it is not $H_1$ but $H_2$ which is
responsible for the possibility of half-integer spin~$^{7}$.  In
fact, the $U(1)$ bundles over $S^2$ are classified by their ``winding
number'' in $H_2(S^2;\Z)$, and the odd winding numbers belong to
spinorial quantum theories.  (Such bundles possess a connection whose
curvature is an odd multiple of the fundamental ``Wess-Zumino term''
on $S^2$).  Now since  copies of $\Z$ in $H_2(....;\Z)$ show up as
copies of $\R$ in $H_2(....;\R)$, equation (11) tells us in this case
that the 2-cycle corresponding to the ``$S^2$ of frames'' remains
nontrivial in $X_{2,1}$;
\[
    1 = dimH_{2}(S^{2};\Reals)  = dimH_{2}(X_{2,1};\Reals).
\]
It follows that the possibility of spinorial states remains as well
(which, by the spin-statistics theorem of Ref.~9, augmented in the
 manner of Ref.~4, is equivalent to the possibility of fermionic
statistics).  Finally, since dim~$H_2(X_{2,1};\R)$ is only 1, we see
that there is no further topologically non-trivial Wess-Zumino term,
beyond the one required for nontrivial spin and statistics.

\section{Outlook}

We have shown that for particles moving in R$^{3}$ and carrying
$SO(3)$-frames, the 1-cycle of exchange (which is homologous to the
1-cycle of rotating a particle frame) is non-trivial in the space
$X_{2,1}$.  We have also obtained analogous results involving $H_2$
for the case of $S^2$-frames.  It remains to extend these
considerations to $X_{3,2}$, and eventually to all the
$X_{m,n}=\overline Q_{m,n}$, and to their union.  One might expect that,
in generalizing the Morse potential
$V$ to $X_{3,2}$ and beyond,   the critical
indices would all enlarge considerably because of the greater
dimensional configuration space associated with the greater number of
particles.  Such an enlargement of the critical indexes would imply
inductively that $dim H_{1}(X_{m,n};K)$ and $dim H_{2}(X_{m,n};K)$
always remain the same as those of $X_{2,1}$, and hence that the first
and second homology groups of $X_{m,n}$ reduce effectively to those of
$S^2$ or $SO(3)$, as the case may be.

To confirm these expectations via direct generalization of our potential
$V$ to arbitrary $X_{m,n}$ would seem difficult, because our analysis of
the critical points in Section 7 was graphical and not analytic.  Instead
of this, one could try to construct the flow $\xi^a$ directly by extending
to arbitrary configurations the more easily defined flow {\it \`a la}
Ref.~11 which retracts a neighborhood of $X_{m+1,n+1}$ back to $X_{m,n}$.
In effect, this is what we have done above for $(m,n)=(1,0)$, and a scheme
for doing something similar in general does not seem too difficult to
devise.

An alternative approach would be to try to generalize the
Mayer-Vietoris techniques of Ref. 11, by means of which results analogous
to those of the present paper were found for $X_{2,1}$ in the rather
simpler, two-dimensional situation.  Such techniques might ultimately
furnish more information on the homology of configuration space, but
in our experience, they have been more cumbersome in application than
the Morse-theory techniques of the present paper, and their complexity
escalates as the spatial dimension grows.

We conclude by returning briefly to the 2-dimensional situation with which
we began Section 8, which concerns particles moving in $\Reals^{2}$ and
carrying $SO(2)$ frames.  In that situation, we showed above that, although
$H_1$ can only decrease in going from $X_{1,0}$ to $X_{2,1}$, $H_2$ can (and
in fact does) increase, giving rise to the possibility of a Wess-Zumino
term in the Action.  In this connection, an interesting question would
arise if the critical indices were indeed to enlarge considerably when
going to $X_{3,2}$ and beyond, as suggested above in the 3-dimensional
case; for this would mean that $H_2$ would necessarily remain nonzero, and
the Wess-Zumino possibility would persist.  We have shown elsewhere$^{11}$
that $H_2(X_{2,1};Z) = Z$ and have exhibited a non-trivial closed two form
defined on $Q_{2,1}$ which vanishes at the lower stratum $Q_{1,0}$.  The
question arises as to what is expected about realizing, say,
$H^{2}(X_{3,2})$ by closed two forms.  Should one expect a closed form on
$Q_{3,2}$ which reduces on approach to $Q_{2,1}$ to the form of
Ref.~11?  We have sought such a form to no avail.  Does this mean that
$H_{2}(X_{3,2}) = 0$?  Or does it mean that $H_{2}(X_{3,2})$ remains
non-zero, but it cannot be realized by closed two forms which go over
continuously from $Q_{3,2}$ to $X_{2,1}$, and if so, then what is the
physical significance of this impossibility?  The answers await further
analysis.

\noindent {\bf Acknowledgments}

\noindent
We would like to thank M. Rothenberg for suggesting the use of
Morse Theory and explaining how it could be applied to our configuration
spaces.
We are also grateful to
A.P.~Balachandran for extensive discussions during the
initiation of this work.  This  work was supported by a
National
Science Foundation cooperative
research grant, INT 8814944.  In addition,
R.D.S. was partially supported by the National
Science Foundation  under grant number PHY 9307570 and by research
funds from Syracuse University and S.S by FORBAIRT. W.D.M. would like to thank
the Dublin Institute for Advanced Studies
for the hospitality shown
during several visits when much of this  work was done, and R.D.S would
like to thank Trinity College Dublin and the Dublin Institute for Advanced
Studies for hospitality during a series of visits.

\pagebreak

\noindent {\bf Figure Captions}
\begin{itemize}
\item Fig.1  Depiction  of subspaces $\Sigma, \Sigma_{t}$, and $M_{t}$ and the
retraction flow
through $x$ .  See text.
\item  Fig.2.   Depiction of subspaces involved when retracting through a
critical point.
See text .
\item  Fig.3 .  Contour plot for $V$ for $\Theta = (\pi,\pi , 0 )$, showing
three critical
points,
one at  $(\theta_{1}, \theta_{2}, \theta , \rho_{1}\ \rho_{2})
\approx (\pi,\pi,0, 1.5, 1.5)$ and a symmetrical pair, one partner of which
is  at
$ (\theta_{1},\theta_{2},\theta,\rho_{1}\ \rho_{2})
\approx (\pi,\pi,0, 0.7, 8.0)$.
\item Fig.4.  Contour plot for $V$ for $\Theta = (\pi,\pi,\pi )$ showing a
critical point at
$(\theta_{1}, \theta_{2}, \theta , \rho_{1}\
\rho_{2})\approx(\pi,\pi,\pi,16,0.6)$.
\item  Fig.5.  Contour plot for $V$ with $\Theta=(0,\pi,\pi)$,
showing a critical point at $(\theta_{1}, \theta_{2}, \theta , \rho_{1}\
\rho_{2}) \approx
(0,\pi,\pi,12,0.6)$.
\item Fig.6.  Geometries of the three configurations which exhibit critical
points.
The heavy arrow represents the antiparticle and the light arrows the particles.
\end{itemize}

\end{document}